# Ionic superlubricity in intercalated bilayers

Zihao Yang, Menghao Wu*

School of Physics and Hubei Key Laboratory of Gravitation and Quantum Physics, Huazhong University of Science and Technology, Wuhan 430074, China

*Email: wmh1987@hust.edu.cn

## Abstract

We propose a type of ionic superlubricity using incommensurate interface to facilitate ion migration, which can be applicable to various layered systems intercalated by ions. For example, our first-principles calculations show that for $NbS_2$ bilayer intercalated by Na ions, the ion migration barriers can be reduced from 0.19 to 0.025 eV upon a large twist angle, and the estimated room-temperature ion conductivity is more than 2 orders of magnitude higher compared with current record of superionicity. The ionic incommensurate interface may also give rise to ion superlubric ferroelectricity as the twisted systems contain various ion coordinations breaking the crystal symmetry, which can be switched via combination of interlayer sliding and ion migration, and the switching barriers reduced by an order of magnitude may enable high speed data writing with low energy cost.

## Introduction

Structural superlubricity utilizing incommensurate contacts in crystalline interfaces to minimize friction becomes an effective lubrication strategy[1-8], which has sparked significant attention in the past decades due its potential to revolutionize various mechanical systems with dramatically enhanced efficiency. According to the Frenkel-Kontorova model[9], when two periodic lattices are misaligned, interlayer static friction can be dramatically suppressed or even eliminated due to the offset of the atomic lateral force, which promises to minimize energy loss, lower maintenance requirements, and extend the operational life of mechanical devices. In 2004, such phenomenon was experimentally confirmed [1], and the limitations of vacuum experimental condition and nanoscale lubricity were overcome by using self-retracting device in a later study [4]. Since then, structural superlubricity has been widely explored both experimentally and theoretically [5, 7, 10-12] in various systems including van der Waals heterobilayers[13-16] (e.g., graphene/$MoS_2$, BN/$MoS_2$, graphene/$SnS_2$, etc.) and twisted homobilayers[17, 18]. Moreover, various disruptive device concepts based on sliding vdW interfaces such as nanobearings without friction and wear, micro-/nanogenerators for self-powering, sliding ferroelectrics, and mechanically reconfigurable devices, have been proposed, leading to the emergence of a new field referred as "slidevices" (5).

In this paper, we try to use similar strategy to minimize ion migration barriers in ion conductors utilizing incommensurate stacking. Solid-state ion conductors exhibiting fast ion transport have been widely used in various energy-storage devices. It is known that the ion conductivity is proportional to $\exp\left(-\frac{\Delta E}{k_B T}\right)$, which can be greatly enhanced even when the barrier $\Delta E$ is only slightly reduced. For the well-known cathode material $LiCoO_2$ and anode material lithiated graphite ($C_6Li$), the migration barriers of Li ions are both around ~0.3-0.4 eV [19-21], leading to low room-temperature ion conductivity around ~$10^{-4}$ S/cm.[22] In such layered ion conductors, uniformly distributed ions occupy identical sites in periodic potentials; consequently, their migration is collectively pinned since a rigid displacement of the ion lattice enforces all ions to cross equivalent barriers along a synchronized pathway. Here we propose that upon a large twist angle making the bilayer stacking incommensurate, the intercalated ions will be located in various local configurations. As shown in Fig. 1(a), the ions are therefore distributed over a spatially heterogeneous landscape, which may substantially lower the average migration barrier. Compared with a few recent predictions of single-ion moiré-potential engineering on intercalated graphene bilayer[23], our proposed collective depinning via ionic superlubricity in a fully intercalated ionic incommensurate interface should be much more effective in barrier reduction.

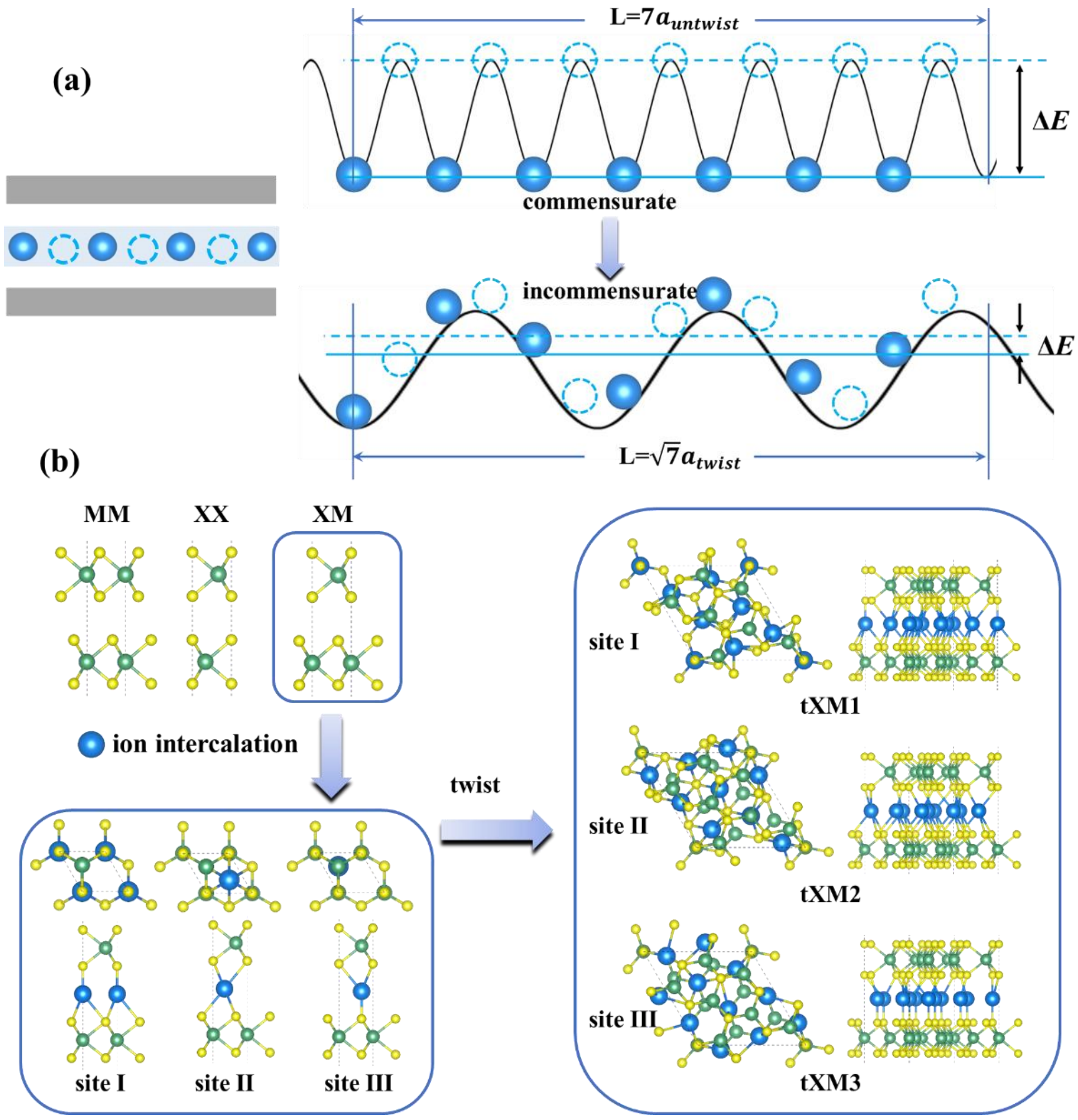

**Fig. 1.** (a) Illustration of ionic superlubricity in bilayer intercalated by ions, where the collective ion migration barrier $\Delta E$ can be greatly reduced as an incommensurate interface is formed by a large twist angle between bilayer. (b) Three configurations of antiparallel stacked bilayer TMDs, where the three sites for ion intercalation of XM, are respectively transformed to tXM1, tXM2 and tXM3 upon a large twist angle.

**Table 1.** The relative energy $\Delta E$ compared with the ground state and polarization $P$ for various configurations of Na-ion intercalated twisted antiparallel bilayer $NbS_2$.

| | tXM1 | tXM2 | tXM3 | tMM1 (=tXX2) | tMM2 (=tXX3) | tMM3 (=tXX1) |
|---|---|---|---|---|---|---|
| $\Delta E$ (meV/f.u.) | 0 | 7.61 | 4.52 | 0.031 | 5.81 | 424.61 |
| $P$ (pC/m) | -0.19 | -0.27 | 0.24 | -0.18 | -0.26 | 0.46 |

**Results and Discussion**

For typical 2H phase transition metal dichalcogenides (TMDs), the three high-symmetry anti-parallel stacking configurations for bilayer are displayed in Fig. 1(b), respectively denoted as XX, MM and MX. Taking the most prevalent XM stacking as an example, there are three ion intercalation sites I, II and III, which are respectively transformed into tXM1, tXM2 and tXM3 when the upper TMD layer is rotated by a large angle that makes the bilayer stacking incommensurate. Similarly, other 6 configurations can be derived from XX and MM upon a large twist angle, where some configurations are exactly equivalent, as listed in Table 1.

To reduce the ion migration barrier utilizing ionic superlubricity, the intercalated ion layer should be rigid in-plane, so the ion radius should be slightly larger than half of the distance between two sites. Here we take Na-intercalated bilayer $NbS_2$ upon a twisted angle of 21.79° as a paradigmatic case for study, noting that due to the challenge for simulating exactly incommensurate interface, large supercells have been adopted for approximation in previous computational studies of structural superlubricity. As listed in Table 1, tXM1 is the ground state, while other phases except for tMM3 (=tXX1) are almost degenerate in energy. Herein tXM2 can be transformed to tMX1 by collective in-plane displacement of Na ions with a low barrier of 0.025 eV/Na (even lower than room-temperature $k_BT$), as shown in Fig. 2(a) and (b). For the long migration of Na ions crossing multiple unitcells (see the pathway tXM1 → tXM2 → tXM1 marked by the color hexagons), the barrier is the same as tXM1-tXM2 transition barrier 0.025 eV/Na, almost an order of magnitude lower compared with the untwisted case (0.19 eV/Na) and lithium ions in graphite (~0.4 eV/Li).

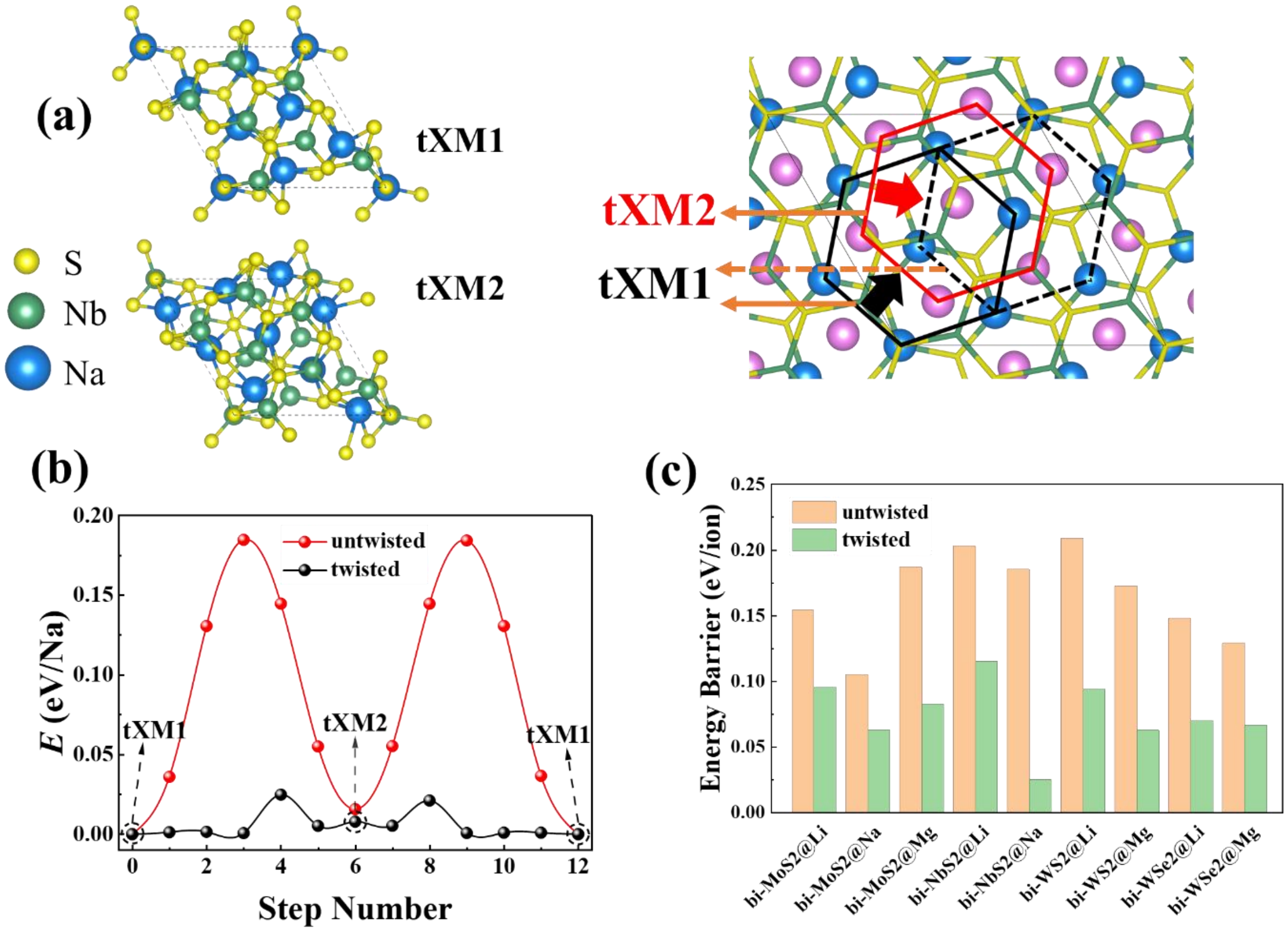

**Fig. 2.** (a) Geometric structures of tXM1, tXM2 and (b) tXM1-tXM2 transition pathway as part of Na migration pathway (black curve) in $NbS_2$ bilayer intercalated by Na ions, much lower compared with the untwisted case (red curve). (c) Ion migration barriers for different intercalated ions (denoted as bi-TMDs@ion) in untwisted and twisted bilayer TMDs.

There is a rough estimation formula for ionic conductivity[24]:

$$\delta_{ion} = (\frac{\gamma c \nu Z^2 e^2 {a_0}^2}{k_B T})\exp(-\frac{\Delta G_m}{k_B T})$$

where $Z$ denotes the charge number of the intercalated ion, while $a_0$ and $\Delta G_m$ represent the migration distance and migration barrier of the intercalated ion, respectively. $c$ refers to the ion density, $\nu = 10^{13}$ $s^{-1}$ is the frequency, and $\gamma = 2$ accounts for the two-dimensional ion diffusion channels. Using these parameters, the room-temperature ionic conductivity $\delta_{ion}$ of Na-ion intercalation in twisted bilayer $NbS_2$ is estimated to be more than 3 orders of magnitude higher compared with the untwisted case. However, this formula was based on single-ion uncorrelated activated-hopping model, while the situations can be much more complicated for collective migration. We further perform molecular dynamics (MD) simulations using Deep Potential machine-learning interatomic potential trained on DFT data to estimate the conductivity, where the relative drift velocity exhibits a clear linear dependence on the electric field (see the details in Supplemental Material Fig. S1-S2). The room-temperature ion conductivity (8.16 S/cm) estimated by such method is 3-4 orders of magnitude higher compared with untwisted case (0.0032 S/cm), also much larger compared with all previously reported superionic conductors [25-28], including the record of 32 mS/cm for inorganic Li-ion conductors[26]. It is noteworthy that compared with real incommensurate interfaces, the migration barriers are actually overestimated as the compensation of lateral force can be influenced by the adopting of large supercells for approximation in our simulations. Hence such ion superlubricity may possess even higher ion conductivity than our predicted values.

Similar strategy in lowering the barrier can be applicable to many other intercalated bilayer TMD systems. Fig. 2(c) summarize the predicted diffusion barriers for various intercalated ions (denoted as bi-TMDs@ion) in various untwisted and twisted bilayer TMDs. In all these cases, the introduction of incommensurate environments by twisting leads to a pronounced reduction in the ion migration barriers, which should result in enhancements by orders of magnitude in ion conductivity.

Such ionic superlubricity can also be used to reduce the switching barriers of intercalation ferroelectricity.[29] Previous studies have shown that ferroelectricity can be induced in van der Waals layers by intercalation of tetra-coordinate metal cations that break the symmetry and give rise to large vertical polarizations,[30, 31] which have been experimentally verified in systems like $CuCrS_2$, $CuCrSe_2$ and $AgCrS_2$.[32-35] However, the switching barriers for reversing such coordination are generally high, e.g., 0.32 eV for $CuCrS_2$, hindering devices from high-speed data writing with low energy cost. Asymmetrical stacking may also break the symmetry in such intercalated layers and give rise to ionic sliding ferroelectricity,[36] where the barriers can be lower.

For non-polar anti-parallel $NbS_2$ bilayer intercalated by Na ions, even the ground state MM-site I is non-polar, some of the configurations like MM-site II, XM-site I and XX-site II, are with nonzero vertical polarizations (see Table 2). For example, due to symmetry breaking, the tetra-coordinate MM-site II and XM-site I exhibit notable differences in the distances between Na layers and two adjacent $NbS_2$ layers (Δd=0.794 and 0.751 Å respectively for MM-site II and XM-site, see Fig. 3(a)), so vertical polarizations of 0.71 pC/m and 0.61 pC/m are respectively formed. However, those polar states are not stable since they are all much higher in energy compared with the ground-state MM-I (see Table 2). For another polar phase XX-II that is only 0.054 eV/f.u. higher in energy, the switching barrier between two opposite polar states may reach 148.7 meV/Na. Herein the ferroelectricity stem from the asymmetric stacking, where Na ions are right over Nb ions at one side, while staggered for the other side, also leading to a small difference in the distances between Na layers and two adjacent $NbS_2$ layers (Δd=0.025 Å).

**Table 2.** The relative energy $\Delta E$ compared with the ground state and polarization $P$ for various configurations of Na-ion intercalated untwisted antiparallel bilayer $NbS_2$.

| | MM- I | MM- II | XM- I | XM- II | XX- I | XX- II |
|---|---|---|---|---|---|---|
| $\Delta E$ (eV/f.u.) | 0 | 0.394 | 0.392 | 0.015 | 1.013 | 0.054 |
| $P$ (pC/m) | non-polar | -0.71 | -0.61 | non-polar | non-polar | 0.07 |

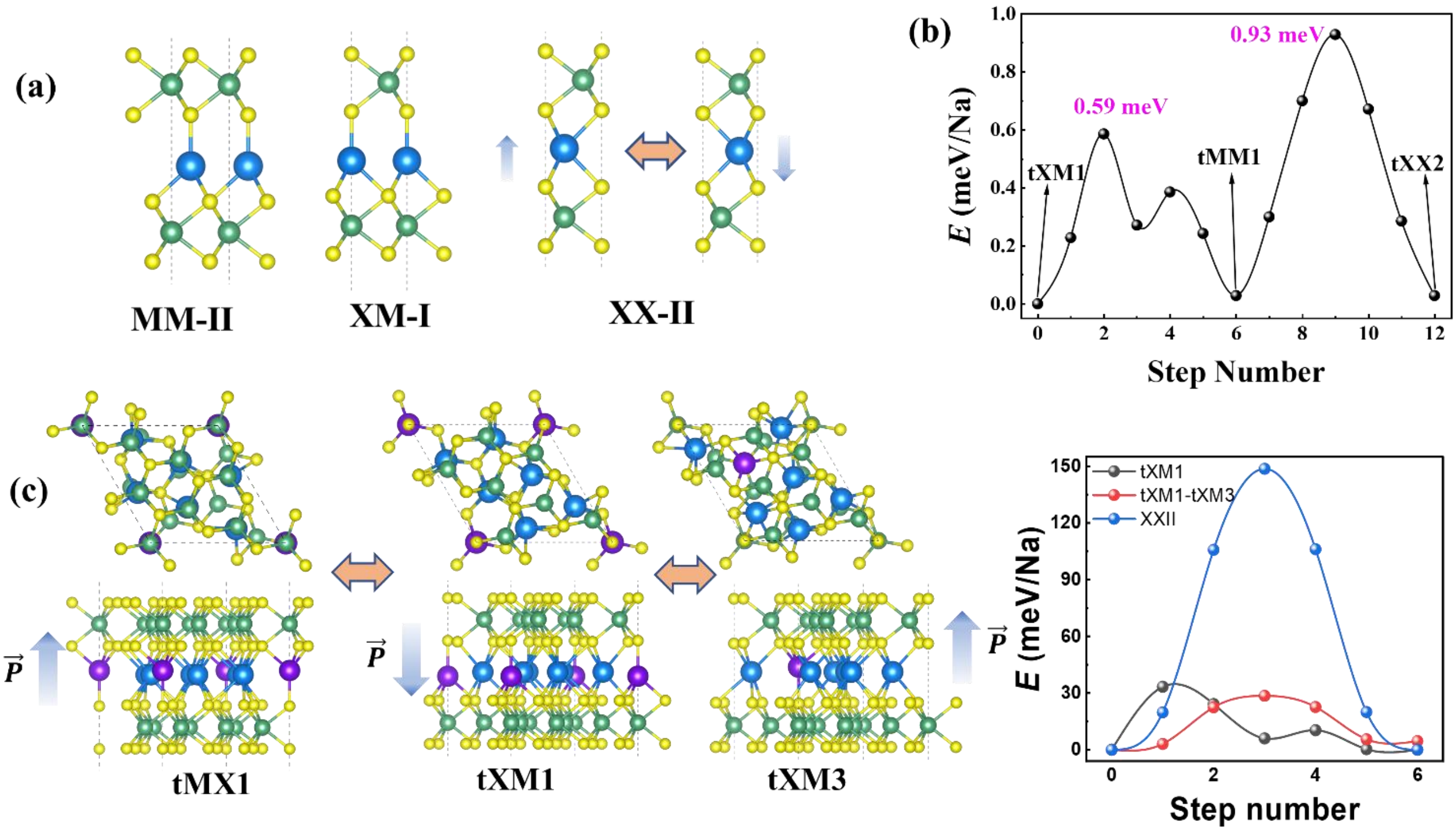

**Fig. 3.** (a) Three high-energy polar configurations of untwisted intercalated bilayer $NbS_2$. (b)Transition pathways between low-energy twisted configurations tXM1→tMM1→tXX2. (c) Ferroelectric switching pathway for tXM1 and tXM1-tXM3 transition pathway, with much reduced barriers compared with ferroelectric switching for untwisted XXII. Herein the Na ions in tetracoordinate configurations are marked in purple.

Upon a large twist angle, the incommensurate intercalated bilayer involving various intercalated sites may render a polar ground state. Actually, all the twisted Na-intercalated configurations in Table 1 possess vertical polarizations ranging from 0.18 to 0.46 pC/m, including the ground state tXM1, tMM1 and tXX2 with 1/7 of intercalated Na ions in tetracoordinate coordination. These nearly degenerate tXM1, tMM1, and tXX2 configurations can be mutually transformed via interlayer sliding, and the barriers for path tXM1→tMM1→tXX2 are respectively 0.59 and 0.93 meV in sequence (see Fig. 3(b)).

The ferroelectric switching of tXM1 involves the interlayer sliding of both $NbS_2$ layers as well as in-plane migration of Na ions, also accompanied by vertical displacements of 1/7 Na ions that are in tetra-coordinate configurations. The switchable polarization is comparable to that of sliding ferroelectric bilayer $MoS_2$,[37] and the switching barrier from tXM1 to the oppositely polarized tMX1 state is 33.3 meV/Na in Fig. 3(c), much reduced compared with untwisted XX-II (148.7 meV/Na), also smaller compared with ionic sliding ferroelectricity in $LiCoO_2$ (71.26 meV).[36] Such ferroelectricity induced by intercalated Na ions in tetracoordinate coordination seems to be robust at 400K as revealed by MD simulations in Fig. S3 (see Supplemental Material). If the two $NbS_2$ layers are fixed, tXM1 with -0.19 pC/m can still be transformed to tXM3 with opposite 0.24 pC/m only via displacements of Na ions, and the switching barrier from tXM1 to tXM3 is further reduced to 28.6 meV/Na.

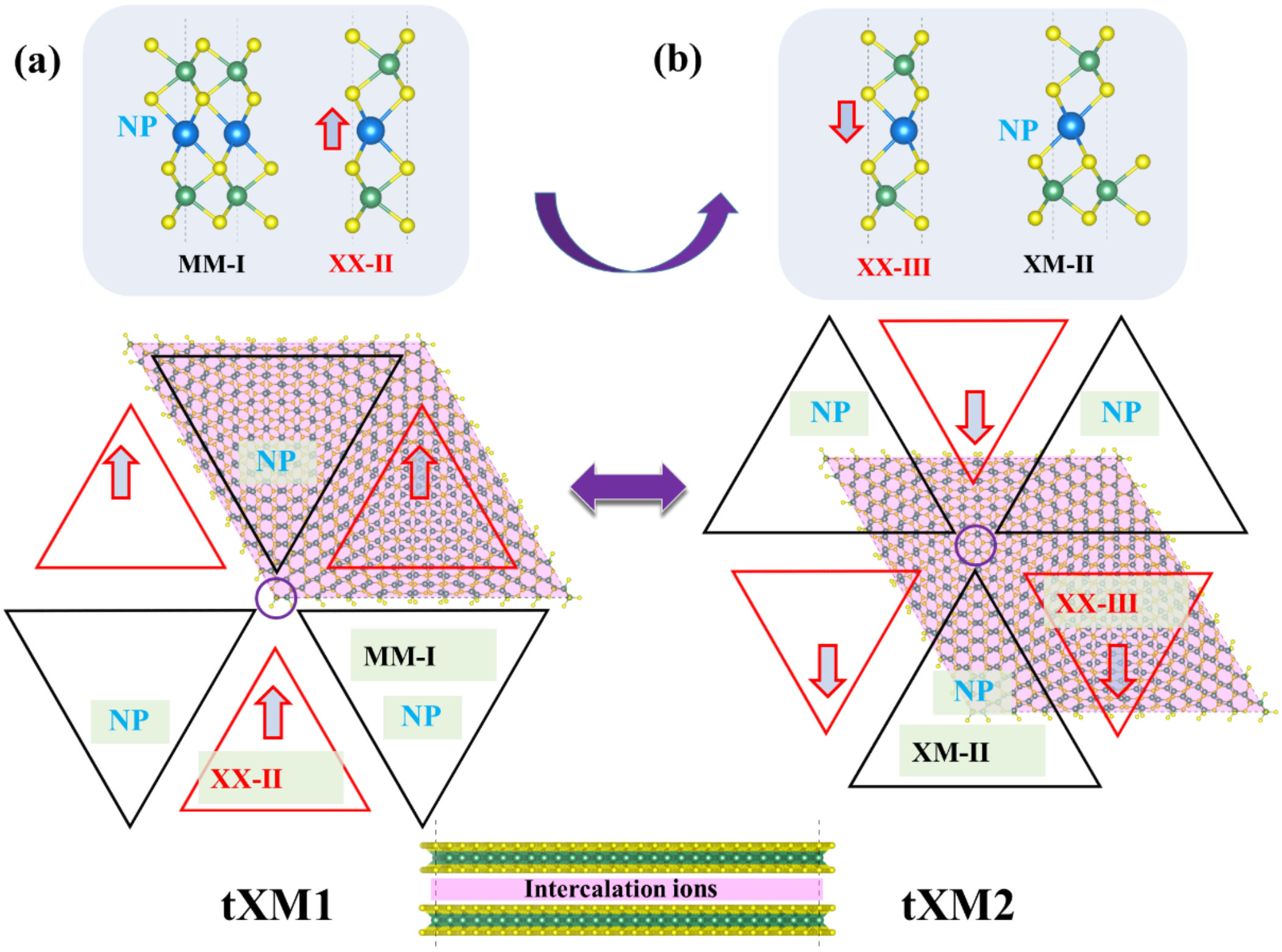

**Fig. 4.** (a) Ionic moire superlattice of tXM1 at small twist angle is mainly composed of nonpolar MM-I and polar XX-II domains. (b) Moire pattern of tXM2 composed of polar XX-III and non-polar XM-II domains, which can be transformed from tXM1.

The systems can still be polar upon a small angle that leads to the formation of a moire superlattice, where the low-energy stacking domains will be expanded while high-energy configuration regions will shrink upon lattice reconstruction. As shown in Fig. 4(a), the reconstructed tXM1 structure is mainly composed of nonpolar MM-I domains and polar XX-II domains. Its net polarization upwards can be reversed by a vertical electric field as it is transformed to the tXM2 in Fig. 4(b) via interlayer sliding + ion migration, where the nonpolar MM-I domains become XX-III domains with polarization downwards, while the polar XX-II domains become non-polar XM-II domains. Such ionic moire ferroelectricity is distinct from previously reported moire ferroelectricity with compensated polarization,[37] which is composed of polar domains of opposite directions.

## Conclusion

In summary, we predict the existence of ionic superlubricity in various layered systems intercalated by ions, and the significantly reduction in ion migration barriers upon twisting have been demonstrated by first-principles calculations. The room-temperature ion conductivity can be enhanced to unprecedented value, orders of magnitude higher compared with current record in superionics. Such ionic incommensurate interface can also be used to induce intercalation ferroelectricity with low switching barriers, where the twisted systems containing various ion coordinations break the inversion symmetry, and the generated vertical polarization can be switched via combination of interlayer sliding and ion migration. Our findings may render technical advances and exotic effects in both fields of superionicity and ferroelectricity.

## Methods

Our first-principles calculations were carried out using the Vienna *Ab initio* Simulation Package (VASP 5.4.4) within density functional theory framework[38, 39]. The projector-augmented wave (PAW)[40] method was used to describe electron-ion interactions, and the exchange-correlation functional was treated by the generalized gradient approximation (GGA) with the Perdew-Burke-Ernzerhof (PBE) parametrization[41, 42]. Brillouin zone sampling was performed using the Monkhorst–Pack scheme[43], with k-point meshes of 21 × 21 × 1 for untwisted systems and 12 × 12 × 1 for twisted systems. A kinetic energy cutoff of 400 eV was applied for the plane-wave basis, and Gaussian smearing with SIGMA = 0.10 eV was adopted. A vacuum spacing of 30 Å along the z direction was introduced to eliminate spurious interlayer coupling. Structural optimizations were performed with convergence thresholds of $10^{-6}$ eV for electronic iterations and 0.01 eV/Å for atomic forces. The climbing-image nudged elastic band (CI-NEB) method was used to calculate the minimum energy migration pathways[44], with the force convergence threshold set to 0.03 eV/Å. The ferroelectric polarization is calculated using the dipole moment correction method [45], where the out-of-plane dipole moment $p_z$ of charge-neutral slab supercell is obtained from the converged total ionic and valence-electron charge distribution:

$$p_z = \int_{\Omega} (z - z_0)\rho_{\mathrm{ions+valence}}(\mathbf{r})\, \mathrm{d}^3 r$$

and herein $z_0$ is the reference center for the dipole moment.

## Supporting Information

## Acknowledgements

We would like to acknowledge the financial support from the National Natural Science Foundation of China (Grant No. 12574263).